\documentclass{ws-procs9x6}

\begin{document}

\title{Cross-over between different symmetries}

\author{ S. FRAUENDORF$^*$ }

\address{Department of Physics, University Notre Dame, IN  46557, USA\\
$^*$E-mail: sfrauend@nd.edu}

\begin{abstract}
The yrast states of even even vibrational and transitional nuclei are interpreted as a rotating condensate
 of interacting d-bosons. The corresponding semi-classical tidal wave concept is used for microscopic
 calculations of energies and E2 transition probabilities.   The strong octupole correlations in the light
rare earth and actinide nuclides are interpreted as rotation-induced condensation of interacting f-bosons. 
\end{abstract}


\bodymatter
\vspace*{1cm}
\section{Cross-over between symmetries} 

\begin{figure}[t]
\begin{minipage}[t]{0.492\linewidth}
\centering
\psfig{file=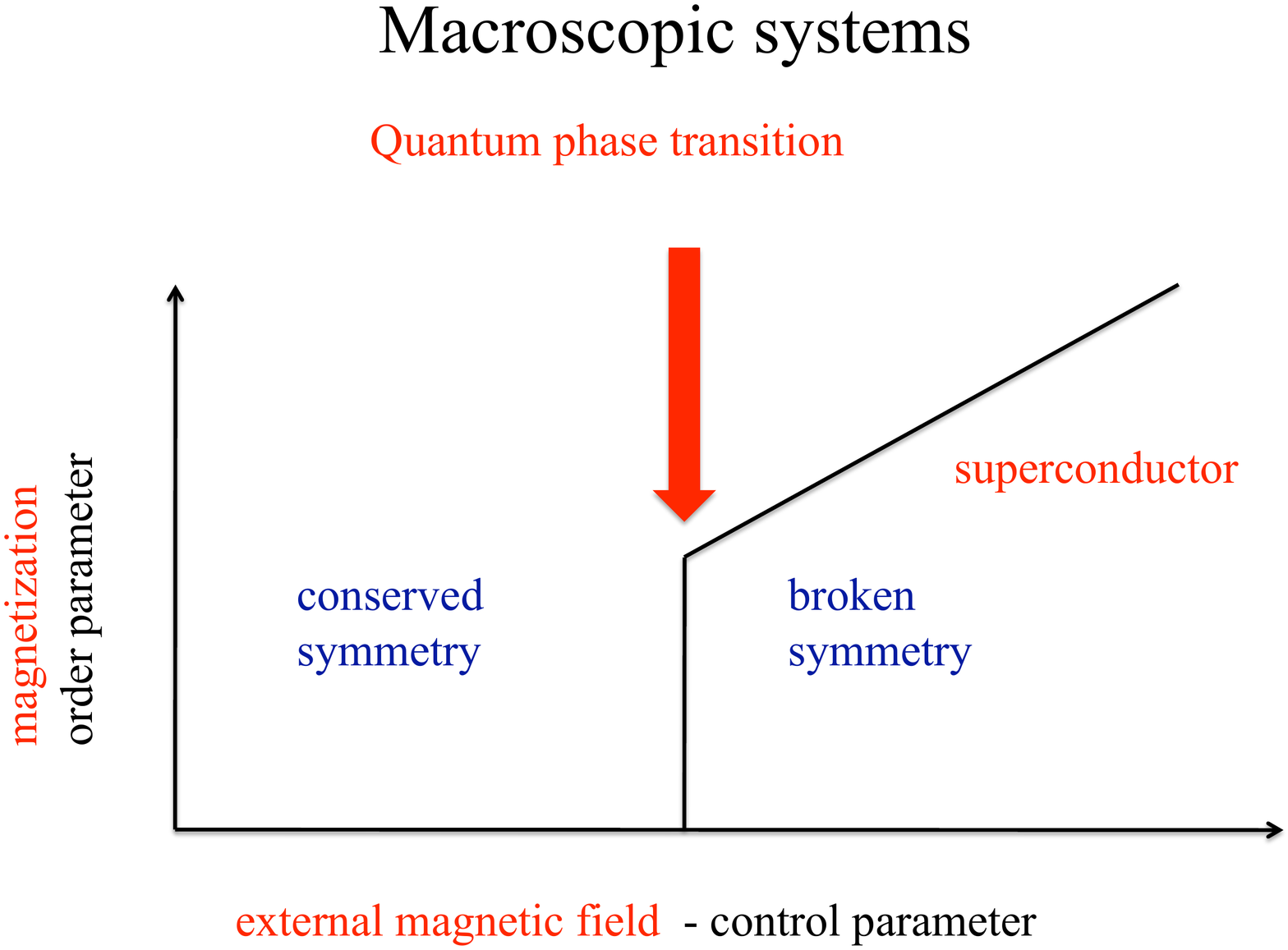,width=1.1\linewidth}
\caption{Quantum phase transition  in a macroscopic
system. Example: Type I superconductor at zero temperature. }
\label{f:macro}       
\end{minipage}
\hfill
\begin{minipage}{0.492\linewidth}
\centering
\vspace*{-3.2cm}
\psfig{file=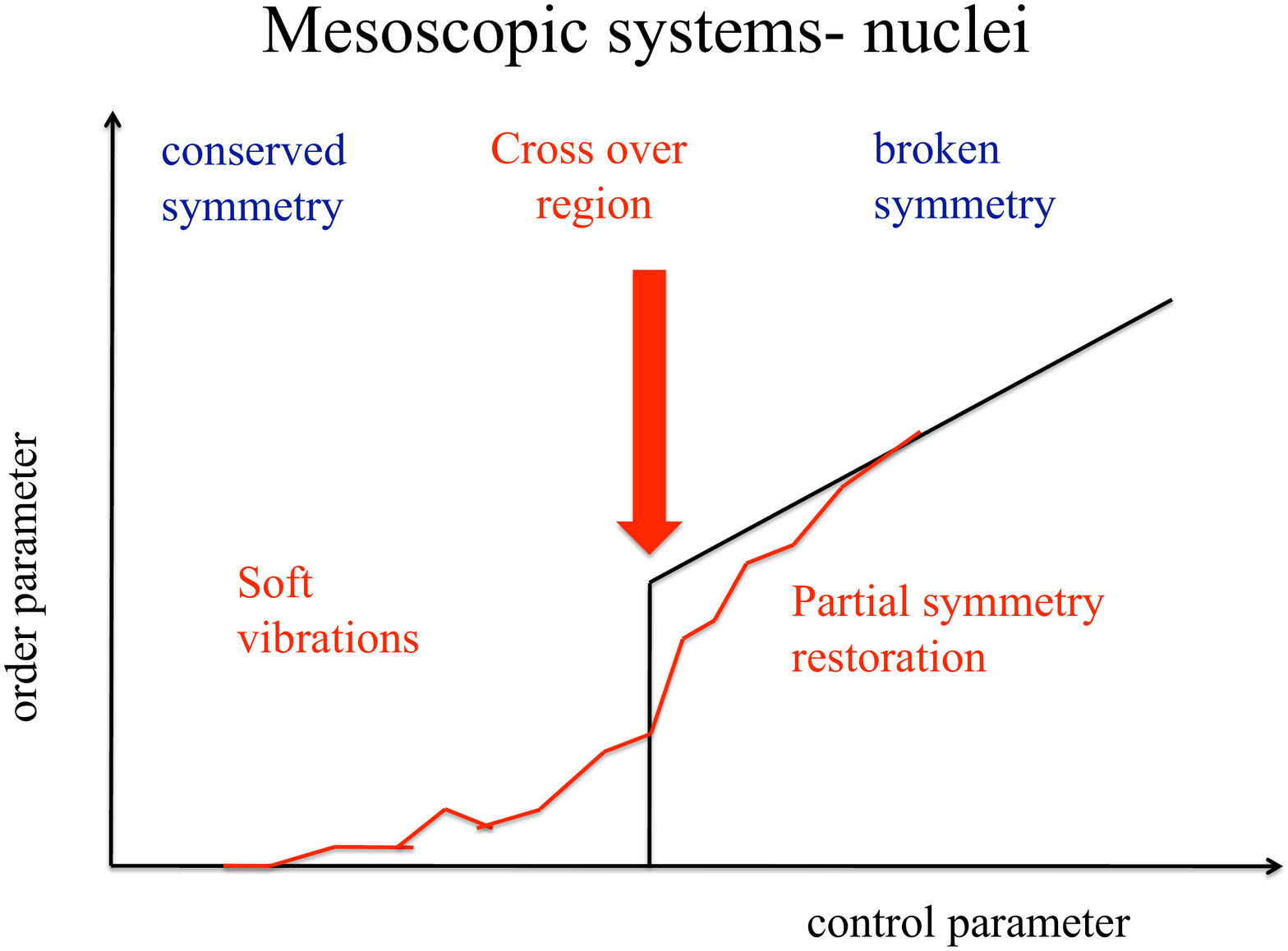,width=1.1\linewidth}
\caption{Appearance of an extended cross-over region in a mesoscopic system}
\label{f:meso} 
\end{minipage}
\end{figure}

In a macroscopic system the transition between distinct symmetries induced by the change of an external
control parameter appears as a phase transition. Fig. \ref{f:macro} schematically illustrates the quantum phase 
transition between the superconducting (gauge symmetry broken)  and normal phases of a type I superconductor.   
In a small mesoscopic system, as the nucleus, the phase transition is not sharp, instead there is a cross-over region.
Starting from the symmetry conserving side, vibrations   dynamically violate the symmetry, which become increasingly
soft and anharmonic.  Starting from the other side, symmetry restoring phenomena, as rotation and tunneling, increasingly
lift the degeneracies induced be the broken symmetry. In the case of nuclei, the control parameters are proton and neutron numbers $Z$, $N$,
and the angular momentum $I$. The symmetries are the ones of the nuclear mean field.  The symmetries of the rotating mean field and 
the ensuing degeneracies have been discussed in Ref. \cite{rmp}. In this talk, I will discuss the cross-over between conserved and broken 
rotational symmetry, as an example for a continuos symmetry, and the cross-over between conserved and broken reflection symmetry, as an 
example for a discrete symmetry. 

\section{Condensation of aligned quadrupole phonons-tidal waves}

\begin{figure}[t]
\begin{minipage}[t]{0.492\linewidth}
\hspace*{-0.3cm}
\psfig{file=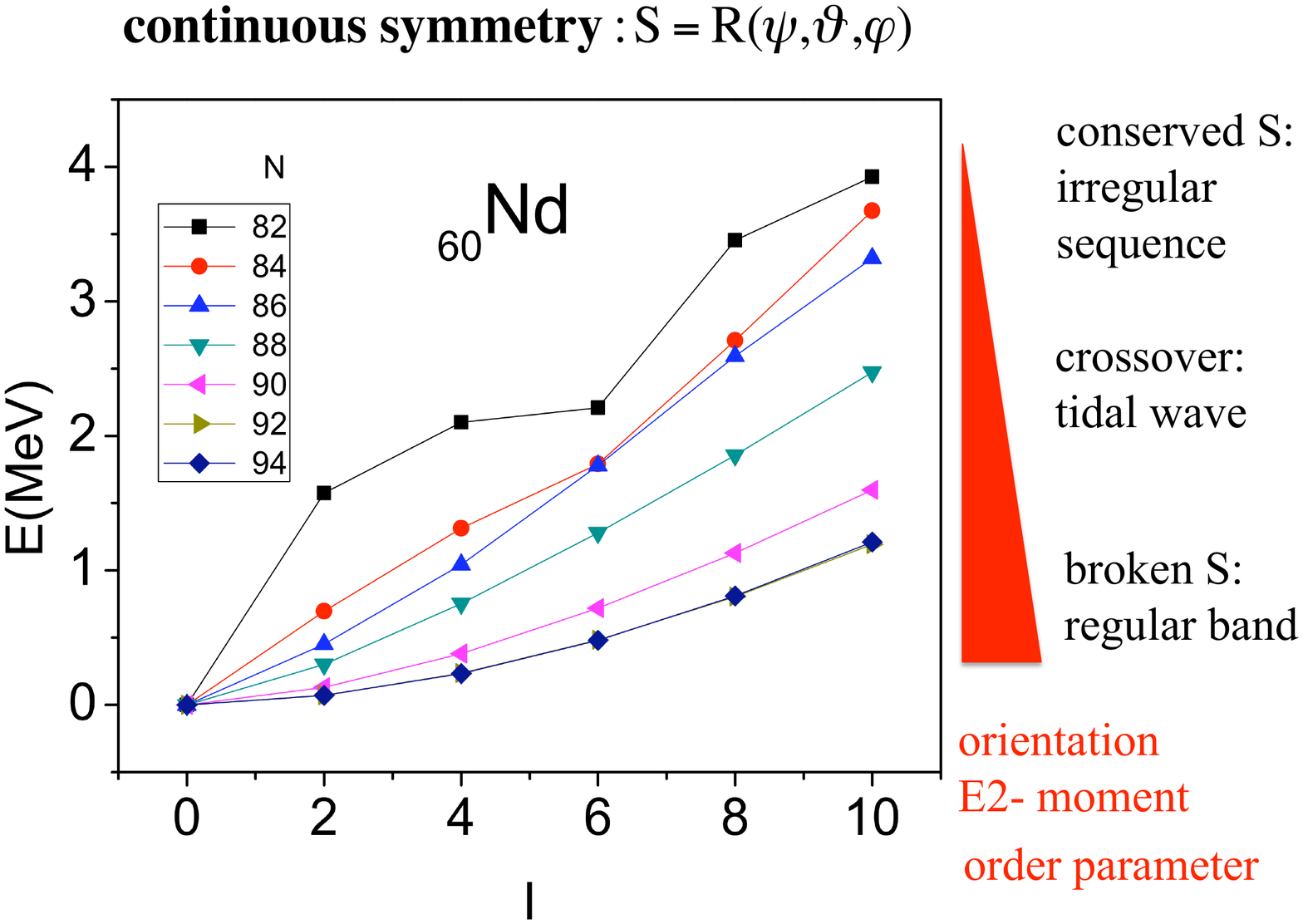,width=1.25\linewidth}
\vspace*{-0.9cm}
\caption{Yrast energies of the Nd-isotopes. The red wedge illustrates the deviation from spherical symmetry.}
\label{f:Ndsys}       
\end{minipage}
\hfill
\begin{minipage}{0.492\linewidth}
\centering
\vspace*{-4.2cm}
\psfig{file=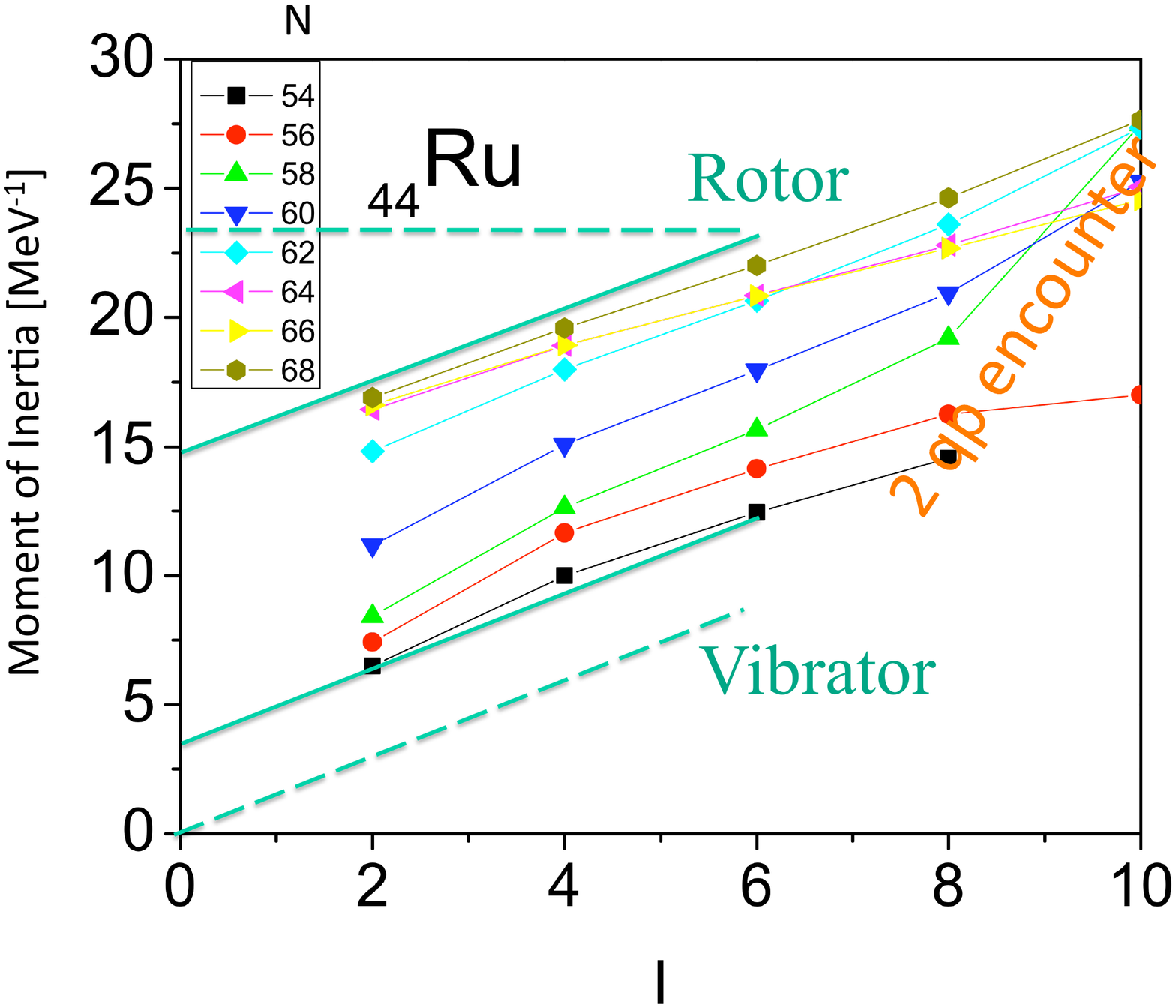,width=1.2\linewidth}
\vspace*{-0.8cm}
\caption{Moments of inertia of the yrast states of the Ru-isotopes}
\label{f:Rusys} 
\end{minipage}
\end{figure}

If the nuclear mean field is spherical, the yrast line is generated  by subsequent 
excitations of particles and holes, where angular momentum is increased by alignment of their 
spins. The spacing between the yrast levels is irregular, reflecting the single particle energies.
If the mean field is deformed, its energy does not depend on its orientation in space, because the Hamiltonian
is rotational invariant.  This spontaneously broken symmetry is restored by the appearance of rotational states, which 
represent superpositions of the symmetry broken (oriented) states. The stronger the orientation (order parameter), 
which is measured by the E2 transition moment, the more rotational ($E= I(I+1)/2{\cal J}$) the yrast line,
 and the larger the moment of inertia ${\cal J}$. Fig. \ref{f:Ndsys} demonstrates 
that there is a gradual development from an irregular yrast sequence  for the semi-magic $N=82$ Nd-isotope
to the rotational sequence of the open shell $N=94$ isotope.  In the center of the crossing-over around $N=86$ 
the yrast energies increase  approximately linearly over an extended range of $I$, which is the
expected soft vibrational sequence. The yrast line is generated by stacking quadrupole phonons,
which align their spins (d-bosons). In other words, there is a condensation of d-bosons. Semi-classically, such a condensate
of aligned quadrupole phonons represents  a quadrupole wave that travels with the angular velocity
$\omega=\Omega_2 /2$ over the spherical nuclear surface, where $\Omega_2 $ is the frequency of the 
quadrupole vibration.  The name "tidal wave" has been suggested \cite{Frau10} because 
 of its similarity with tidal waves on the ocean.

 The surface of a tidal wave moves  with the constant angular velocity $\omega$ as the one of a rotor.
 However there is a difference. The energy and 
 the angular momentum increase with the amplitude of the harmonic tidal wave wave, whereas the frequency stays constant.
 The energy and the angular momentum of the rigid rotor increase with the angular frequency while
 the shape remains unchanged. The difference is manifest by the moment of inertia ${\cal J}(I)=I/\omega$, which is proportional
 to $I$ for the tidal wave and constant for the rotor.  Fig. \ref{f:Rusys} shows the experimental function ${\cal J}(I)=2I/(E(I)-E(I-2))$ 
 the Ru-isotopes. The experiment lies between the limiting cases of a harmonic vibrator (tidal wave) and a rigid rotor.   
As expected, there is a gradual  transition from the vibrational behavior for $N=54$ (center of the cross-over) to rotational 
one for $N=64$ (broken symmetry). 

\begin{figure}[t]
\begin{center}
\vspace*{-1cm}
\psfig{file=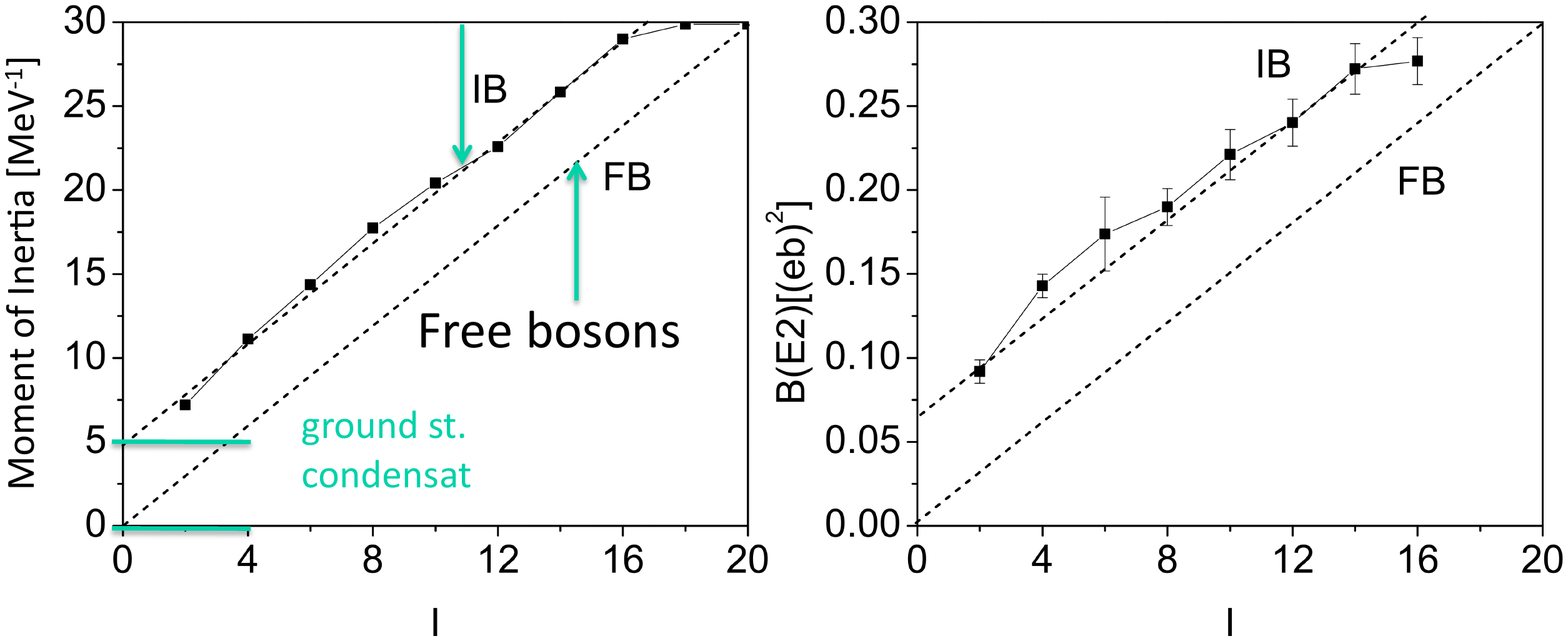,width=\linewidth}  
 \end{center}
 \vspace*{-3.5cm}
 \caption{The moment of inertia $J$ (left) and the  $B(E2,I\rightarrow I-2)$ transition probabilities (right)
of the yrast states of $^{102}$Pd. The dashed line FB (free bosons)
shows the limit of harmonic bosons. The dashed line IB (interacting bosons) illustrates the near linear
trend of the interacting bosons. }
\label{f:JIBE2}
\end{figure}

The interpretation of the near-equidistant yrast sequences as d-boson condensation implies that
the reduced transition probability $B(E2, I\rightarrow I-2)$ should approximately linearly increase with $I$.
%The identification of complete vibrational multiplets is troubled by the fact that  
% with increasing phonon number $n$, the collective states are embedded into a progressively dense background of quasiparticle excitations.
 % The coupling  to the quasiparticle background 
%fragments the collective states, which cease to exist as individual quantum states. The density of quasiparticle excitations 
%is lowest near the yrast line, which is the sequence of states with minimal energy for a given angular momentum $I$. 
%With increasing phonon number $n$,
%the yrast members of the vibrational multiplets keep their identity as collective quantum states longest. 
%Moreover, they couple to high-j two-quasiparticle excitations, which have a simple structure.  
%Hence, the  vibrational states with highest
%phonon number are expected at the yrast line. In this talk, we discuss $^{102}_{46}$Pd$_{56}$. 
The recent lifetime measurements by A. D. Ayangeakaa {\it et al.} \cite{lifetimes},  
have confirmed the expected increase  in $^{102}$Pd up to the seven-boson yrast state case. 
The results are displayed in Fig. \ref{f:JIBE2}. The $B(E2)$ values increase with $I$ in the same way as
 the moment of inertia, such that their ratio $B(E2)/{\cal J}$ is independent of $I$ within the experimental uncertainties. 
This clearly demonstrates that the yrast line is generated by stacking d-bosons or,   using the semi-classical interpretation,
by increasing the amplitude of the tidal wave.
 
Fig. \ref{f:JIBE2} shows that the yrast line of $^{102}$Pd classifies as an anharmonic tidal wave. 
The moment of inertia is a nearly linear function of $I$ indicated by the line IB (interacting bosons).
 It deviates from the harmonic limit FB (free bosons) by the small offset at $I=0$, which is a measure of the anharmonicity. 
 The $B(E2)$ values behave in the same way, such that
the ratio $B(E2)/{\cal J}$ is constant. The Liquid Drop Model \cite{BMII}
  suggests that
both $BE(2)$ and ${\cal J}$ are $\propto \beta^2$, which implies that their ratio is $I$-independent. 
  The offset at $I=0$ indicates
 that the ground state must have some deformation due to fluctuations that are larger than
 the zero point fluctuations of the harmonic vibrator.
In classical terms, the tidal wave starts with a small deformation, which  increases along the 
yrast line.   In quantum language, the condensate of interacting bosons rotates
like a  condensate of free aligned d-bosons to which a small fraction of non-aligned d-bosons is added.

\begin{figure}[t]
\begin{center}
\vspace*{-1cm}
\psfig{file=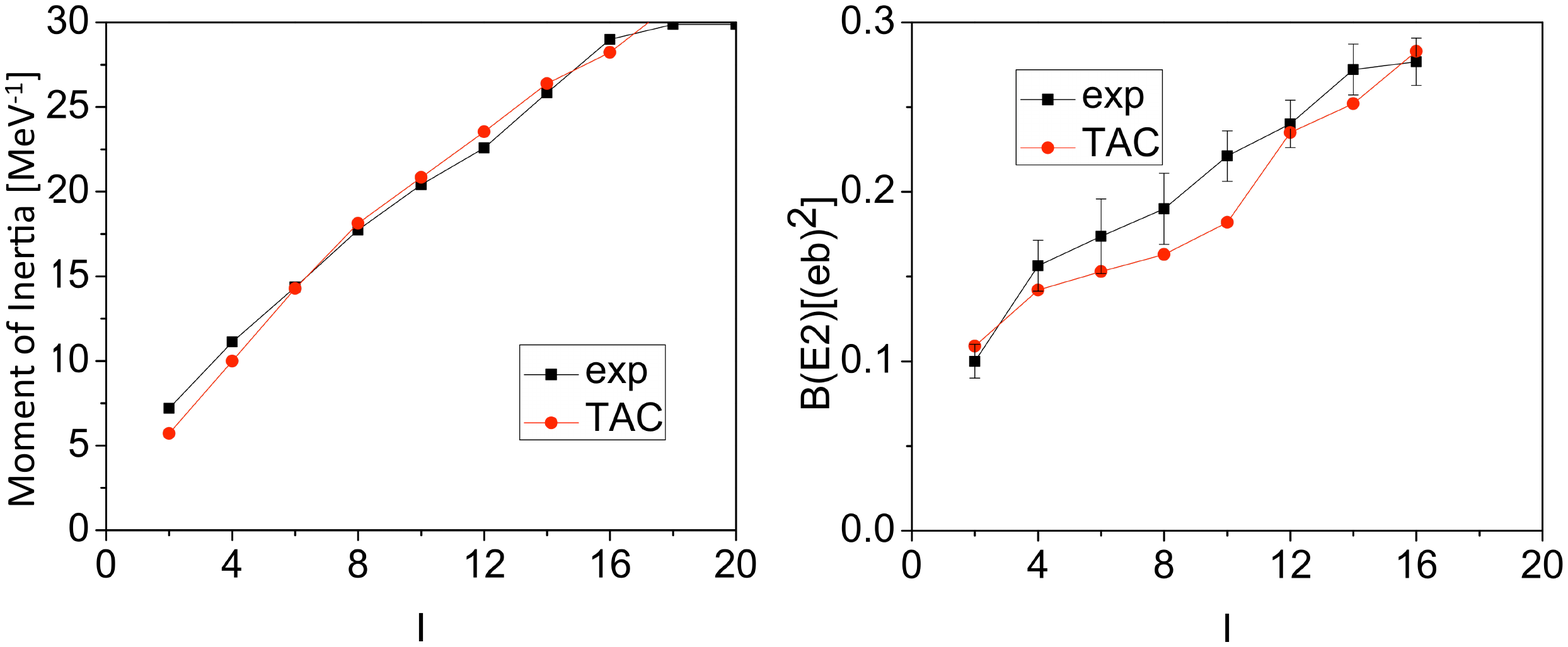,width=\linewidth}  
 \end{center}
  \vspace*{-3.5cm}
 \caption{The moment of inertia ${\cal J}$ (left) and the  $B(E2,I\rightarrow I-2)$ transition probabilities (right)
of the yrast states of $^{102}$Pd compared with the microscopic  calculations, which are  obtained by minimizing
the SCTAC energy (standard LDM, WS potential, and paring) with respect to the deformation parameters $\beta$ and $\gamma$.   }
\label{f:JIBE2TAC}
\end{figure}

The tidal wave has a static deformed shape in the co-rotating frame of reference. This has lead to the microscopic
description suggested by Frauendorf, Gu, and Sun \cite{Frau10}, which is based on the rotating mean field.
We used the   SCTAC (shell correction tilted axis cranking) version \cite{qptac} of 
the Cranking Model, which calculates the energy
for a given expectation value of the angular momentum operator  equal to $I$  by means of the micro-macro method 
using a deformed Woods-Saxon potential. The energy is minimized with respect to the deformation 
parameters $\beta$ and $\gamma$.  \footnote{There are certain technical problems finding the cranking solution
in the near-vibrational regime, which are discussed in Ref. \cite{Frau10}.} Fig. \ref{f:JIBE2TAC} shows that 
in the case of $^{102}$Pd, 
the function ${\cal J}(I)$ is very well reproduced. The calculated $B(E2)$ values show the characteristic increase with  $I$.
They fluctuate stronger than the experimental values, which is most likely due to the neglected zero point fluctuations of the shape.
Note that there are no free parameters adjusted to the experiment. 
We  applied the same method 
to the even-even nuclides with $44\le Z\le 48$ and $56\le N\le 66$ \cite{Frau10}. 
Deformed solutions were found for $I\ge 2$ even when 
the solution was spherical for $I=0$.
These calculations describe the collective yrast states rather well. They also 
describe the intrusion of the aligned h$_{11/2}$ two quasi neutron states into the yrast line, which 
causes the back bending phenomenon seen in most of the studied nuclei.

\section{Condensation of aligned octupole phonons}

\begin{figure}[t]
\begin{center}
\vspace*{-1cm}
\psfig{file=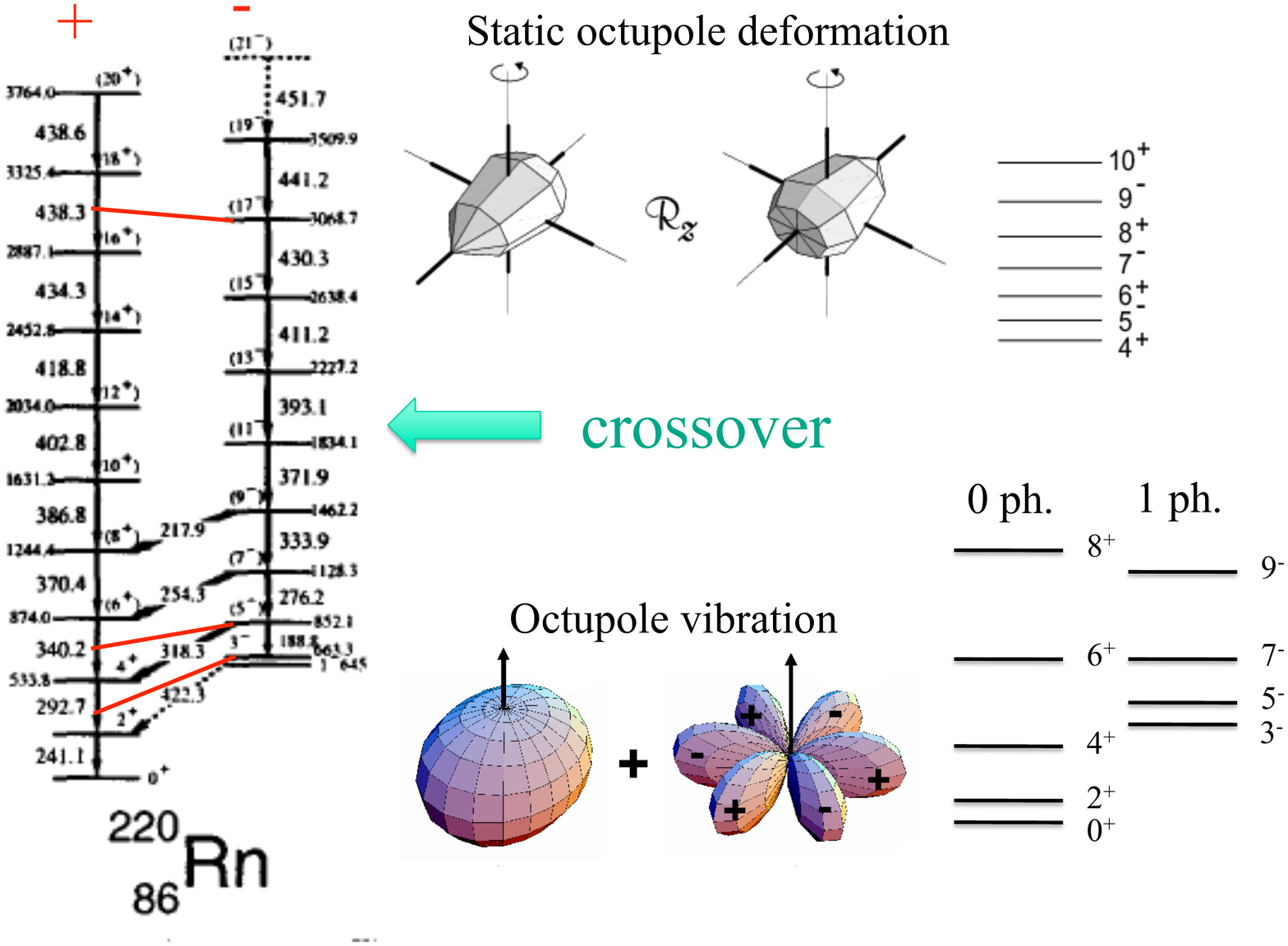,width=\linewidth}  
 \end{center}
 \caption{Octupole vibrations, static octupole deformation as compared to experiment }
\label{f:octcross}
\end{figure}

So far I have discussed quadrupole deformation, which is reflection symmetric, and the 
case that the angular momentum is perpendicular to one of the reflection planes. This symmetry
implies that the collective sequences  (vibrational or rotational bands) have a definite parity and 
$I$ increases in steps of 2, i. e. $I=\alpha+2n$, where $\alpha$ is the signature quantum number of the intrinsic state.    
If the deformed nuclear shape breaks reflection symmetry such that it  still contains two reflection planes
and the angular momentum is perpendicular to one of them,  then the bands are composed
of states with alternating parity, the sequence of which is determined by the simplex quantum number $s$
of the intrinsic state, such that $\pi=s\exp(-i\pi I)$ \cite{BMII}. Fig. \ref{f:octcross} compares  the sequence expected for $s=1$ 
with the  experimental yrast sequences of both parities $^{220}$Rn. For  $I = 10 - 15$ one observes  the expected
interleaving  of states of opposite parity. However, the $\pi=-$ sequence is higher than
 the $\pi=+$ sequence for $I<10$ , and  the $\pi=+$ sequence is higher for $I>15$.  The Figure also shows the two yrast sequences
 of a reflection symmetric nucleus that is soft with respect to octupole deformation. In this case the $\pi=-$ yrast 
 sequence is generated by exciting an octupole phonon that aligns its spin of 3 with the total  angular momentum
 (f-boson). Obviously, the low-spin part of the experimental spectrum looks more like the zero and one-phonon
 bands than the  alternating parity sequence of a static octupole shape. In the remainder, I will suggest that
 the  known "octupole deformed" nuclei are situated in the cross-over region between reflection symmetry and 
 asymmetry, where most of them behave like soft anharmonic vibrators.  

\begin{figure}[t]
\begin{center}
\vspace*{-0.5cm}
\psfig{file=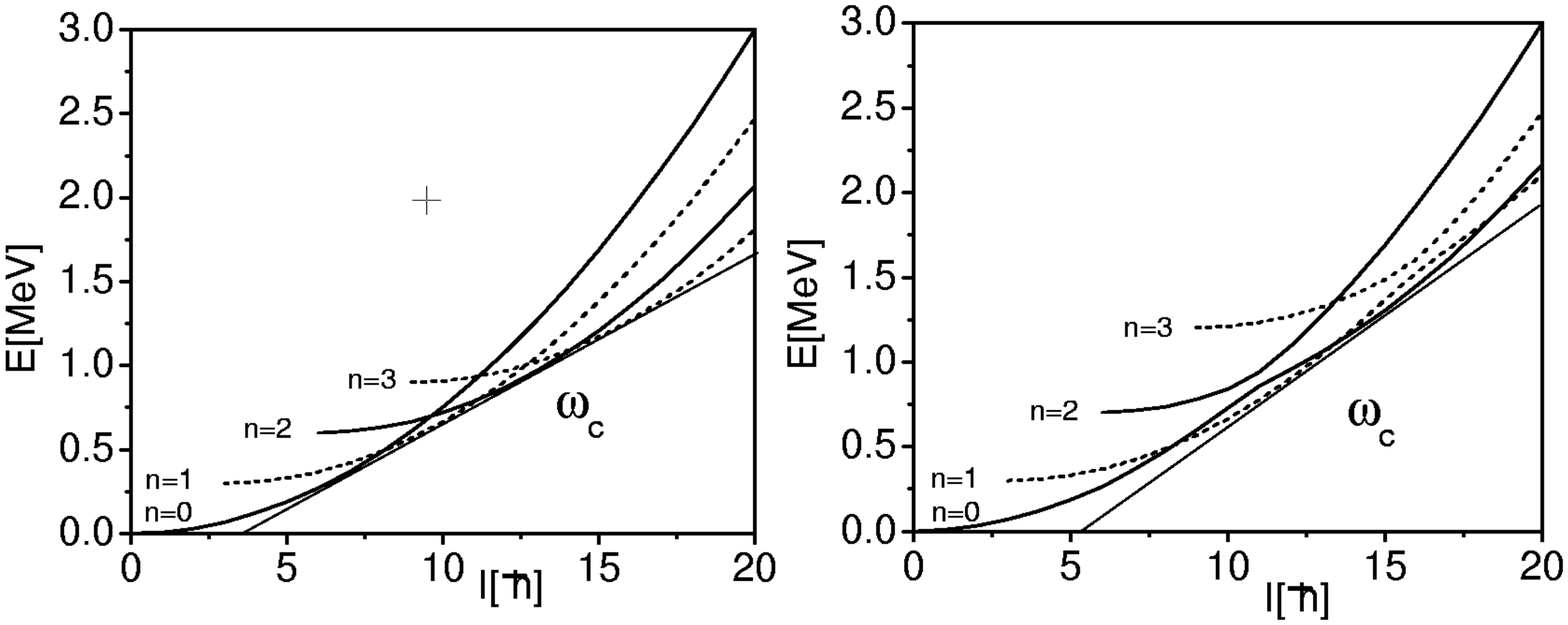,width=\linewidth}  
 \end{center}
  \vspace*{-4cm}
 \caption{Energies of aligned octupole phonon bands. Left: harmonic phonons. Right: anharmonic, interacting phonons.   
 The phonon number is $n$. }
\label{f:condens}
\end{figure}

Fig. \ref{f:condens} illustrates the behavior in cross-over region, which I have discussed in Ref. \cite{octcond}.  
In the left panel it is assumed that the octupole vibrations are harmonic and decoupled from the reflection symmetric  rotor.   
The octupole phonons carry 3$\hbar$ of angular momentum, which aligns   with the rotational axis.
Above the critical frequency $\omega_c$ it becomes energetically favorable to excite a phonon instead of further
increasing the angular velocity of the rotor, which decreases because the phonon adds 3$\hbar$ of angular momentum.
This process is repeated, resulting in a rotation-induced condensation of octupole phonons. Since the octupole phonons 
have negative parity, the $\pi=+$ and $\pi=-$ yrast sequences change  order with their subsequent excitations, which 
is a hallmark of the condensation.   The multi phonon states decay by $E1$ transitions. The dipole transition operator 
arises from coupling the isovector $E3$ moment of the phonons with the isocalar $E2$ moment of the rotor. As a consequence,  
 only the transitions \mbox{$(n,I)\rightarrow (n-1,I-1)$} are allowed. The transitions \mbox{$(n,I)\rightarrow (n+1,I-1)$} are 
 forbidden, because they 
 correspond to a change of rotor angular momentum of 4$\hbar$, which cannot be facilitated by the $E2$ moment. 

\begin{figure}[t]
\begin{center}
\vspace*{-1cm}
\psfig{file=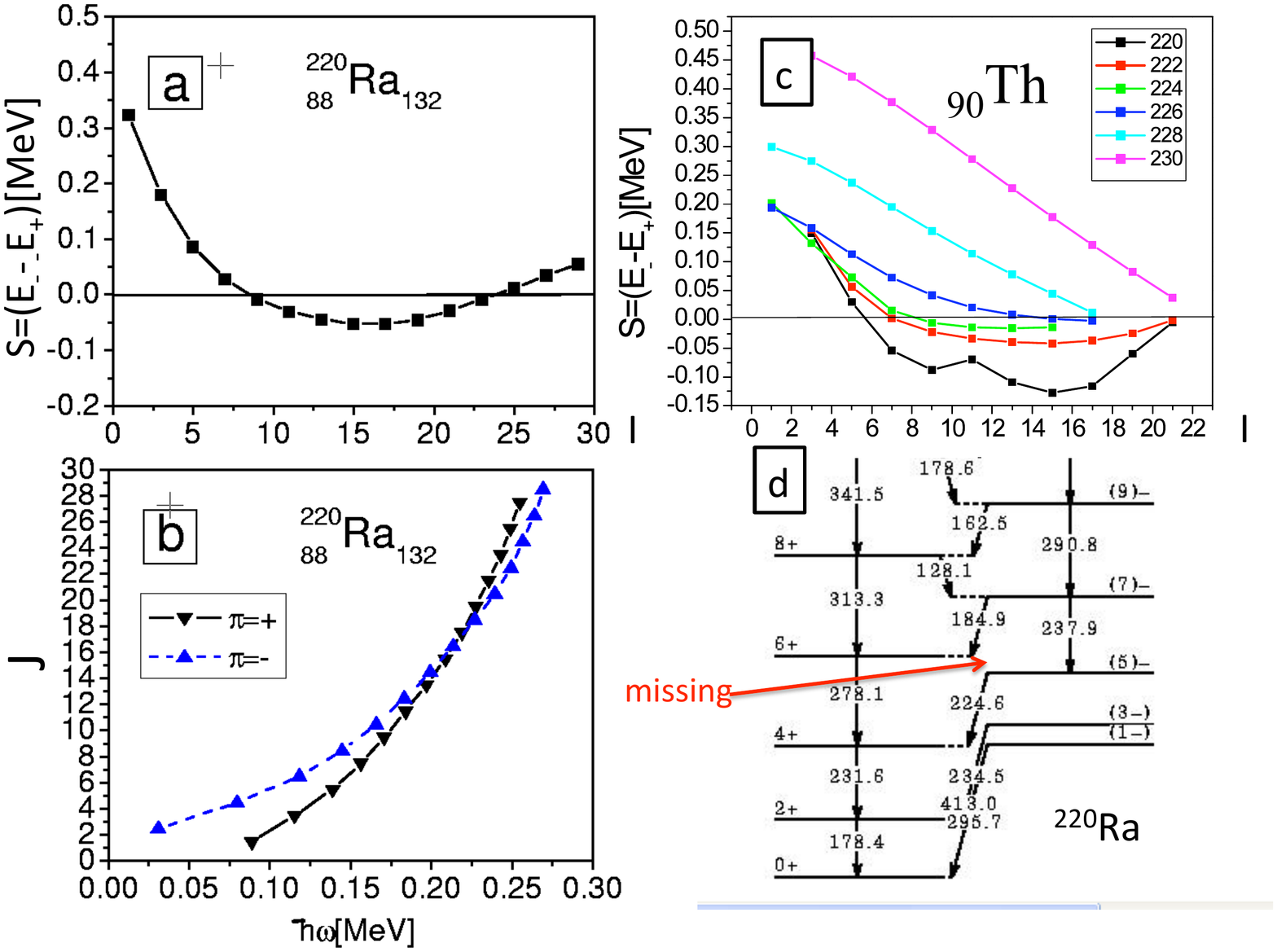,width=\linewidth}  
 \end{center}
 \caption{ a, b, c: Experimental energy difference $S=E_--E_+$ and  angular momentum $J(\omega)$ of the $\pi=-$ and $\pi=+$ yrast
 sequences in $^{220}$Ra and the Th isotopes.  d: Experimental decay scheme of $^{220}$Ra.}
\label{f:RaTh}
\end{figure}

The phonons strongly interact  in the cross-over region, which causes a repulsion between crossing bands of the same parity.
As shown in the right part of Fig. \ref{f:condens},
the sharp crossings of the one-phonon with the zero- and two-phonon bands remain,
 because they have opposite parity. However, the zero- and two-phonon 
 bands mix and repel each other at the avoided crossing. The same holds  for the one- and three-phonon bands. 
 As a consequence, the energy difference $S=E_--E_+$ between
between the yrast sequences of both parities oscillates as function of $I$.
The experimental energy difference $S(I)=E_-(I)-(E_+(I-1)+E_+(I+1))$  for $^{220}$Ra
in Fig. \ref{f:RaTh} a)  display this characteristic  pattern. The one-phonon
band crosses the zero-phonon band before it feels much of the
two-phonon band. At the crossing, $S$ changes sign. 
When the zero-phonon band encounters the tow-phonon one, the two
states mix and exchange character (avoided crossing). The
level repulsion attenuates the growth of $-S$, which starts
decreasing when the $\pi= +$ band has become predominantly
the two-phonon state. When the two-phonon band crosses
the one-phonon band, $S$ changes sign again. Its growth is
attenuated and reversed when the avoided crossing between the
one- and three-phonon bands is encountered, the beginning
of which is still visible.
The angular momentum functions $J_\pm(\omega)$ in  Fig. \ref{f:RaTh} b)
reflect the condensation as follows.   The
 $\pi= -$ one-phonon band starts with additional 3$\hbar$  relative to the  $\pi= +$
zero-phonon band at the same $\omega$, which is the expected
angular momentum carried by an aligned octupole phonon.
The difference decreases, when the  $\pi= +$ two-phonon band,
which carries additional 6$\hbar$, starts mixing into the zero-phonon
band. The  $\pi= -$ and $\pi= +$ bands have equal
angular momentum at the frequency of maximal mixing. Near the one-two-phonon band
crossing at $I = 24$, where the mixing is small, the angular
momentum difference is -3$\hbar$. Fig. \ref{f:RaTh} d) shows that 
the forbidden  $E1$ transition $+\rightarrow -$  is indeed missing at low $I$, where the
one- and zero-phonon bands are still rather pure. At larger $I$,  the transitions
  $+\rightarrow -$ appear because the two-phonon, from which the transition to the one-phonon
   band is allowed, starts mixing in.

The  aligned octupole phonon is a wave  that runs with the angular velocity $\omega_3=\Omega_{oct}/3$ over the nuclear surface.
The rotor rotates with the angular velocity of $\omega_2$. If $\omega_2=\omega_3$ the two motions combine to the
rotation of a reflection asymmetric shape (c.f. Fig. \ref{f:octcross}). The difference $\omega_--\omega_+$  between the slopes of the two yrast sequences reflects
the difference between the angular velocities of the octuople wave and the rotor.  The interaction between the octupole phonons,
which causes the repulsion between the crossing bands, makes the $\pi=\pm$ yrast sequences more parallel. It  tends to lock the two
types of motion to the rotation of a stable reflection asymmetric shape, which is non-axial (heart-shaped) in contrast to the conventionally  discussed
axial octupole deformation (pear-shaped) \cite{Butler}.  Fig. \ref{f:RaTh} c) shows $S(I)$ for the light  Th isotopes.  The $N=134$ and 136 isotopes come closest
to the case of static octupole deformation.  Mean field calculations predict the most stable octupole deformation for these neutron numbers \cite {Butler}. 
 
\begin{figure}[t]
\begin{center}
\vspace*{-1cm}
\psfig{file=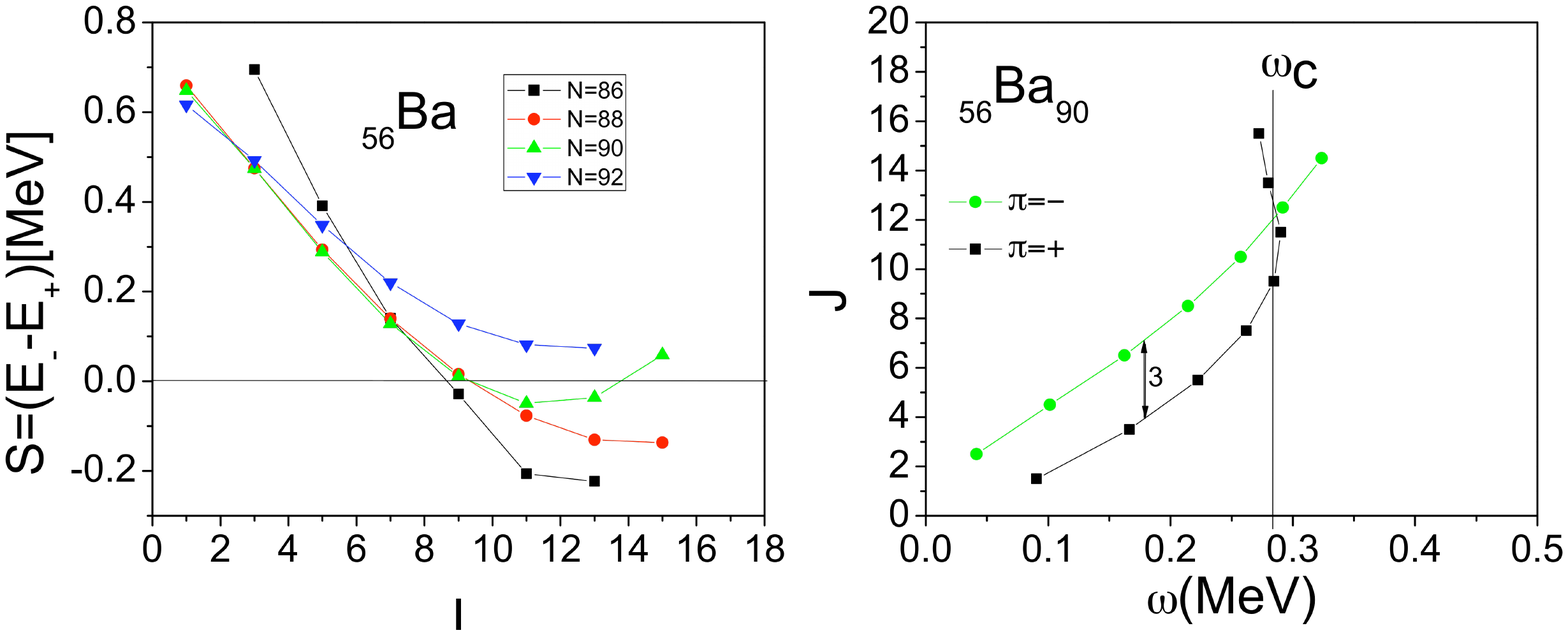,width=\linewidth}  
 \end{center}
  \vspace*{-3.5cm}
  \caption{Experimental energy difference $S=E_--E_+$ and  angular momentum $J(\omega)$ of the $\pi=-$ and $\pi=+$ yrast
 sequences in the Ba isotopes.}
\label{f:Baoct}
\end{figure}

The other light actinides behave similar to the Th isotopes. 
Fig. \ref{f:Baoct} shows the $Z=54$ Ba isotopes as an example from the lighter mass region of strong octupole correlations (c.f.\cite{Butler}).
The cross-over pattern is clearly seen.  As far as studied well enough, the  $Z=54,$ 58, 60, and 62 isotones behave in a similar way.
 For $Z\geq 64$ or $N\geq 94$ the collectivity of the octupole phonons gets lost.  The discussed nuclides with strong octupole correlations
 are not good rotors. They are situated in region of the cross-over toward stable quadrupole deformation, discussed in the preceding section.
 My  discussion assumed a rotor for simplicity. It applies to the tidal waves as well, provided they are anharmonic, such that    
there is an increase of the angular velocity with $I$. Then the multi-octupole phonon bands  are generated by stacking aligned quadrupole phonons
onto aligned octupole phonons. % \footnote{  
The rotation-induced condensation of octupole phonons has been also observed in $^{240}$Pu, which is a good rotor \cite{240Pu}. 
%}  

Alternatively, one may interpret the yrast region in terms of a reflection-asymmetric tidal wave that travels  over the nuclear surface. This
explains why our calculations in  the framework of {\em reflection symmetric} tidal waves did not well well describe the 
$Z=$58 - 62, $N=$ 84 - 90 region. The restoration of parity by tunneling between the shape and its mirror image will cause the splitting between
the two yrast sequences of opposite parity. However, the repeated interchange of their order is not obvious from this standpoint. 
In order to describe the phenomenon in a microscopic way, one should carry out parity-projected cranking calculations of a non-axial,
 reflection-asymmetric mean field, which is rather challenging.  In collaboration with F. D\"onau, we have started developing a 
 microscopic approach from the vibrational side.
 The octupole phonons are described by means of the Tamm-Dankoff method in the rotating frame of reference 
 (in oder to avoid the problems of the RPA near the instability). The interaction
 between the phonons is calculated and a diagonalization with in a space of few phonon excitations will be carried out.     
 
 \section{Summary} 
The yrast states of transitional nuclei  are interpreted in terms of a  rotating d-boson condensate, which 
corresponds   a tidal wave running over the nuclear surface. The seven-phonon yrast state has been
 identified in $^{102}$Pd.    
The boson interaction generates anharmonicity, which 
shows up as a constant shift of the $B(E2,I\rightarrow I-2)$ value and the moment of inertia $J(I)$ as functions of the spin.  
The semi-classical tidal wave concept allowed us carrying out  microscopic calculations based on the Cranking Model,  which 
 reproduce the energies the $B(E2)$ values without adjustable parameters.  All nuclei showing strong octupole 
 correlations are in the cross-over region between reflection symmetry and asymmetry. The properties of their  yrast states are more
 adequately  described 
 by the excitation of rotational aligned octupole phonons, which strongly interact, than by a reflection-asymmetric rotor.       

\section*{Acknowledgments}
Supported by the DoE Grant DE-FG02-95ER4093. I thank A. D. Ayangeakaa and U. Garg for making their lifetime data available to me.

\bibliographystyle{ws-procs9x6}
\bibliography{ws-pro-sample}

\end{document}